\documentclass[aps,prl,twocolumn,superscriptaddress,showpacs,amsmath,amssymb]{revtex4-1}
\usepackage{amsmath}
\usepackage{amsfonts}
\usepackage{amssymb}
\usepackage{graphicx}
\usepackage{color}

\usepackage{bm}
\usepackage{braket}
\usepackage{hyperref}

\begin{document}

\title{Thermodynamic signature of the $\mathrm{SU}(4)$ spin-orbital liquid and symmetry fractionalization from the Lieb-Schultz-Mattis theorem}

\author{Masahiko G. Yamada}
\email[]{myamada@mp.es.osaka-u.ac.jp}
\affiliation{Department of Materials Engineering Science, Osaka University, Toyonaka 560-8531, Japan}
\author{Satoshi Fujimoto}
\affiliation{Department of Materials Engineering Science, Osaka University, Toyonaka 560-8531, Japan}
\affiliation{Center for Quantum Information and Quantum Biology, Osaka University, Toyonaka 560-8531, Japan}

\date{\today}

\begin{abstract}
The $\mathrm{SU}(4)$ Heisenberg model on the honeycomb lattice
is expected to host a quantum spin-orbital liquid at low temperature
with an astonishing candidate material, $\alpha$-ZrCl$_3$.
We employed the canonical thermal pure quantum state method to
investigate the finite-temperature phase of this model.
Exploiting the full symmetry of $\mathrm{SU}(4)$, the calculation
up to a 24-site cluster, which is equivalent to 48 sites in the spin-1/2
language, is possible.  This state-of-the-art computation with large-scale
parallelization enables us to capture the thermodynamic properties
of the $\mathrm{SU}(4)$ Heisenberg model on the honeycomb lattice.
In particular, the specific heat shows a characteristic peak-and-shoulder
structure, which should be related to the nature of the low-temperature
quantum spin-orbital liquid phase.  We also discuss what can be concluded
from the assumption that the ground state is gapped and symmetric
in view of the generalized Lieb-Schultz-Mattis theorem.
\end{abstract}

\maketitle

\textit{Introduction}.---
Quantum spin liquids are an unusual state of matter without a long range
order beyond the Landau paradigm~\cite{Balents2010,Savary2017}.  Specifically,
Kitaev materials, such as $d^5$ iridates and $\alpha$-RuCl$_3$, have
attracted attention because of its explicit
fractionalization~\cite{Kitaev2006,Jackeli2009,Plumb2014,Yamada2017,Yamada2017xsl,Liu2018,Sano2018,Kasahara2018,Jang2019,Yokoi2020,Tanaka2020}.
As a consequence, the thermodynamic signatures including
specific heat and thermal Hall conductivity show a characteristic
behavior~\cite{Nasu2014,Nasu2017}, which is difficult to explain
without rewriting spin operators with Majorana fermions and is
therefore regarded as a precursor of fractionalized topological excitations.

However, the Kitaev model does not have a continuous symmetry,
differently from usual $\mathrm{SU}(2)$-symmetric spin liquids, and
the signature of fractionalization must also be different.
For example, the two-peak structure of specific heat was discussed
in the kagome spin liquid~\cite{Elser1989}, but later the low-temperature
peak has been shown to disappear in large-scale thermal pure quantum (TPQ)
state calculations~\cite{Sugiura2012,Sugiura2013,Hyuga2014},
and now it is not regarded as indication of fractionalization in the
kagome spin liquid.

In spin liquids with a global symmetry $G$, a concept called symmetry
fractionalization plays an important role.  Low-energy excitations
called anyons generically have a $G$-charge which is described by
a projective representation of $G$~\cite{Kitaev2006}.  This
means that the symmetry action gets nonlocal in space, which should
be one of the most definite signatures of quantum spin liquids.

A completely different avenue of physics of symmetry fractionalization
appears in $d^1$ systems with a strong spin-orbit
coupling~\cite{Yamada2018,Natori2018,Yamada2021su4},
which seemingly have a lower symmetry in the spin space.
However, an emergent $\mathrm{SU}(4)$ symmetry appears in $d^1$
systems, which would lead to multicomponent frustration between spin
and orbital degrees of freedom~\cite{Corboz2011,Corboz2012,Lajko2013,Keselman2020,XPYao2021csl,Jin2021,XPYao2021}.
In this case, as is exemplified in honeycomb $\alpha$-ZrCl$_3$,
an emergent $\mathrm{SU}(4)$ symmetry may lead to a rich
symmetry fractionalization which is not expected in a usual
$\mathrm{SU}(2)$ systems.

Then, how can we prove the existence of symmetry fractionalization?
Is there any thermodynamic signature?  To see this, it is important
to define each thermodynamic quantity for each symmetry sector.
For example, the calculation of the specific heat defined for
each symmetry sector essentially requires the full treatment
of the $\mathrm{SU}(4)$ symmetry in finite-temperature
simulations, which is challenging because the dimension of
the Hilbert space gets larger.

Therefore, we developed a state-of-the-art canonical thermal
pure quantum (cTPQ) state method~\cite{Sugiura2013} to solve this problem.
Exploiting the full symmetry of $\mathrm{SU}(4)$ enables us
to calculate a specific heat up to a 24-site cluster, which
is equivalent to 48 sites in the spin-1/2 language.  This is
comparable to the Hilbert space size achieved by the cutting-edge
ground state exact diagonalization method~\cite{Lauchli2019}.
We stress again that our calculation is finite-temperature,
which is enabled by a large-scale parallelization in supercomputers
and a graphics processing unit (GPU).

Although the ground state of the $\mathrm{SU}(4)$ Heisenberg model
on the honeycomb lattice was proposed to be a gapless Dirac spin
liquid in the previous study~\cite{Corboz2012}, we modestly propose
another scenario, a possibility of a gapped symmetric ground state.
In this gapped spin-orbital liquid case, the fractionalization of
the $\mathrm{PSU}(4)$ symmetry can be proven from the refined version
of the generalized Lieb-Schultz-Mattis theorem~\cite{Lieb1961,Affleck1986,Yamada2018,Yamada2021su4}.
In this sense, the symmetry fractionalization is more visible in gapped
spin-orbital liquid, which may be detected by some thermodynamic signatures.

In this Letter, we theoretically present the finite-temperature
specific heat of the $\mathrm{SU}(4)$ Heisenberg model on the
honeycomb lattice.  The specific heat shows a characteristic
peak-and-shoulder structure, which would be useful to identify
the realization of this model in real materials.  The results
are consistent with a gapped scenario, and we also discuss
what can be concluded from the assumption that the ground state
is gapped and symmetric.

\textit{Model}.---
The $\mathrm{SU}(4)$ Heisenberg model on the honeycomb lattice
is defined as follows.
\begin{equation}
    H = \sum_{\langle ij \rangle} \left(2\bm{S}_i\cdot \bm{S}_j+\frac{1}{2}\right)\left(2\bm{T}_i\cdot \bm{T}_j+\frac{1}{2}\right),
\end{equation}
where $\bm{S}_i$ are spin-1/2 operators for the spin sector,
and $\bm{T}_i$ are spin-1/2 operators for the orbital sector.
$\langle ij \rangle$ runs over every nearest-neighbor bond of
the honeycomb lattice.  We put a fundamental representation of
$\mathrm{SU}(4)$ per site, which consists of 2 spin and 2 orbital
degrees of freedom.  It is better to rewrite the Hamiltonian by
the swapping operator.
\begin{equation}
    H = \sum_{\langle ij \rangle} P_{ij},
\end{equation}
where $P_{ij}$ is a swapping operators for two fundamental
representations on the $i$th and $j$th sites of the honeycomb lattice.
In this way, the $\mathrm{SU}(4)$ symmetry is made explicit.
This is a natural generalization of the $\mathrm{SU}(2)$ Heisenberg
model to $\mathrm{SU}(4)$ and can be realized, for example, as
a low-energy effective model of $\alpha$-ZrCl$_3$~\cite{Yamada2018}.
From now on we consider an $N$-site cluster of the honeycomb lattice
with a periodic boundary condition.

\textit{Method}.---
In order to compute finite-temperature quantities for the above model,
we employ the cTPQ method (the Hams-de Raedt method~\cite{Hams2000}).
In the finite-size system, the cTPQ method is regarded as a stochastic
approximation of a trace of a large matrix.  To compute a physical
observable $A$, we use
\begin{equation}
    \mathrm{Tr}\, A e^{-\beta H}\sim \braket{0|e^{-\beta H/2}A e^{-\beta H/2}|0},
\end{equation}
where $\ket{0}$ is a Haar random vector in the Hilbert space.
In practice, it is better to expand the $e^{-\beta H/2}$ by a power
of $H$, and we define
\begin{equation}
    \ket{k} = (l-H)^k\ket{0},
\end{equation}
where $l$ is a real value larger than the maximal eigenvalue of $H$.
$e^{-\beta H/2}$ is expanded until the $\Lambda$th power of $l-H$.
With a large $N$ limit, a single random vector $\ket{0}$ is enough
to compute any physical quantities, but in practice it is better
to sample $N_\textrm{sample}$ vectors and take an average.

For simplicity, we set $l=\max_i\{E_i\} + 1$, $\Lambda=2000$,
where $E_i$ is each eigenvalue of $H$.
For calculations with $N \leq 20$, we also set $N_\textrm{sample} = 100$.
Errors are estimated by the jackknife method.  For $N=24$,
it is difficult to take an average, so we show raw data for
two samples (shown as \#1 and \#2).

In reality, we use the above cTPQ method for each symmetry sector
of $\mathrm{SU}(4)$ and sum the results up afterwards because
the Hamiltonian is block-diagonal.  This direct-sum decomposition
is achieved by the method developed for exact diagonalization
by Nataf and Mila~\cite{Nataf2014,Nataf2016}.  However, their original method
is not suitable for a large-scale parallelization, so we modify
their idea a little to improve the computational efficiency.

\begin{figure}
    \centering
    \includegraphics[width=8.6cm]{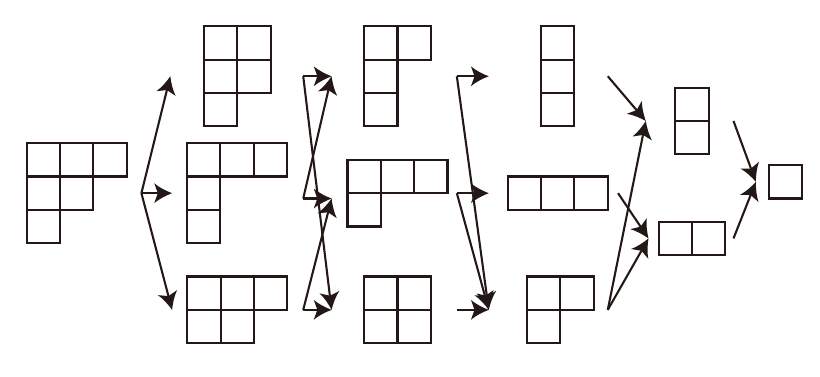}
    \caption{A graph associated with a Young tableau (3, 2, 1) in
    the Wilf-Rao-Shanker method.  Arrows are connected when
    a reduced Young tableau is obtained by eliminating a corner box
    of the original Young tableau.  Each path starting from (3, 2, 1)
    to (1) corresponds to a single SYT by assigning a number from
    1 to 6 to an added box when going backwards from (1) to (3, 2, 1).}
    \label{wrs}
\end{figure}

As is usually the case, the difficulty in the exact diagonalization
or the cTPQ method arises when we make a lookup dictionary for the
subspace basis.  In case of Nataf and Mila's method~\cite{Nataf2014},
such a problem is reduced to the problem to index and retrieve
standard Young tableaux (SYTx) because each basis of a specific
symmetry sector associated with a Young tableau $\gamma$ (with $N$ boxes)
is labelled by an SYT with the same shape as $\gamma$.  We note that
SYTx are prepared by filling $\gamma$ with numbers, 1, 2, \dots, $N$
with a constraint that they are aligning in the ascending order both
in the row and the column.  The problem of indexing and retrieving
such SYTx for a fixed $\gamma$ is solved by Wilf, Rao, and
Shanker~\cite{Wilf1977,Rao2015}. We note
that a similar idea was discussed in Ref.~\cite{Scharfenberger2012}.
In their method, a graph is associated with each $\gamma$ by
eliminating a corner box recursively from a Young tableau $\gamma$
(see Fig.~\ref{wrs}).  Then, we can identify each SYT with each
path starting from $\gamma$ to the a single box.  These paths
are indexed and retrieved by Wilf's method~\cite{Wilf1977} for
any of such graphs very efficiently, and by this method we can
parallelize the exact diagonalization or the cTPQ method for
any $\mathrm{SU}(N_c)$ Heisenberg models.

For calculations with $N \leq 20$, we use a GPU machine
with a CUDA implementation.  For $N=24$, we use a large-scale
supercomputer with a message-passing-based cTPQ implementation.
The flat MPI parallelization up to 18,432 processes is used.
The multiplication of offdiagonal components is possible through
MPI Alltoallv. All codes are written in the Julia
language~\cite{Julia2017,Besard2018,Byrne2021}.

\begin{figure}
    \centering
    \includegraphics[width=8.6cm]{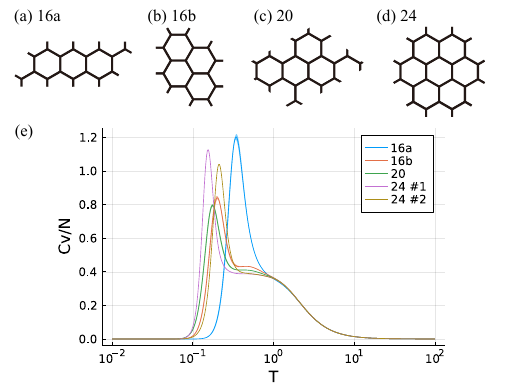}
    \caption{(a) Shape of the 16a cluster.  (b) Shape of the 16b cluster.
    (c) Shape of the 20-site cluster. (d) Shape of the 24-site cluster.
    (e) Temperature dependence of the specific heat for the above
    clusters calculated by the cTPQ method.  For $N \leq 20$-site
    clusters we take an average over $N_\textrm{sample} = 100$ samples
    and errors are shown in ribbons,
    while for the 24-site cluster we show raw data for the two samples
    (shown as \#1 and \#2).}
    \label{main}
\end{figure}

\textit{Specific heat}.---
The specific heat shows a peak-and-shoulder structure for all the system sizes
calculated (see Fig.~\ref{main}).  This typical shape becomes clearer
in the larger system size.  Physically the low-temperature peak is
associated with a color gap, \textit{i.e.} an energy scale of
excitations with an $\mathrm{SU}(4)$ color charge, and the
high-temperature shoulder is associated with the interaction energy scale.

\begin{figure}
    \centering
    \includegraphics[width=8.6cm]{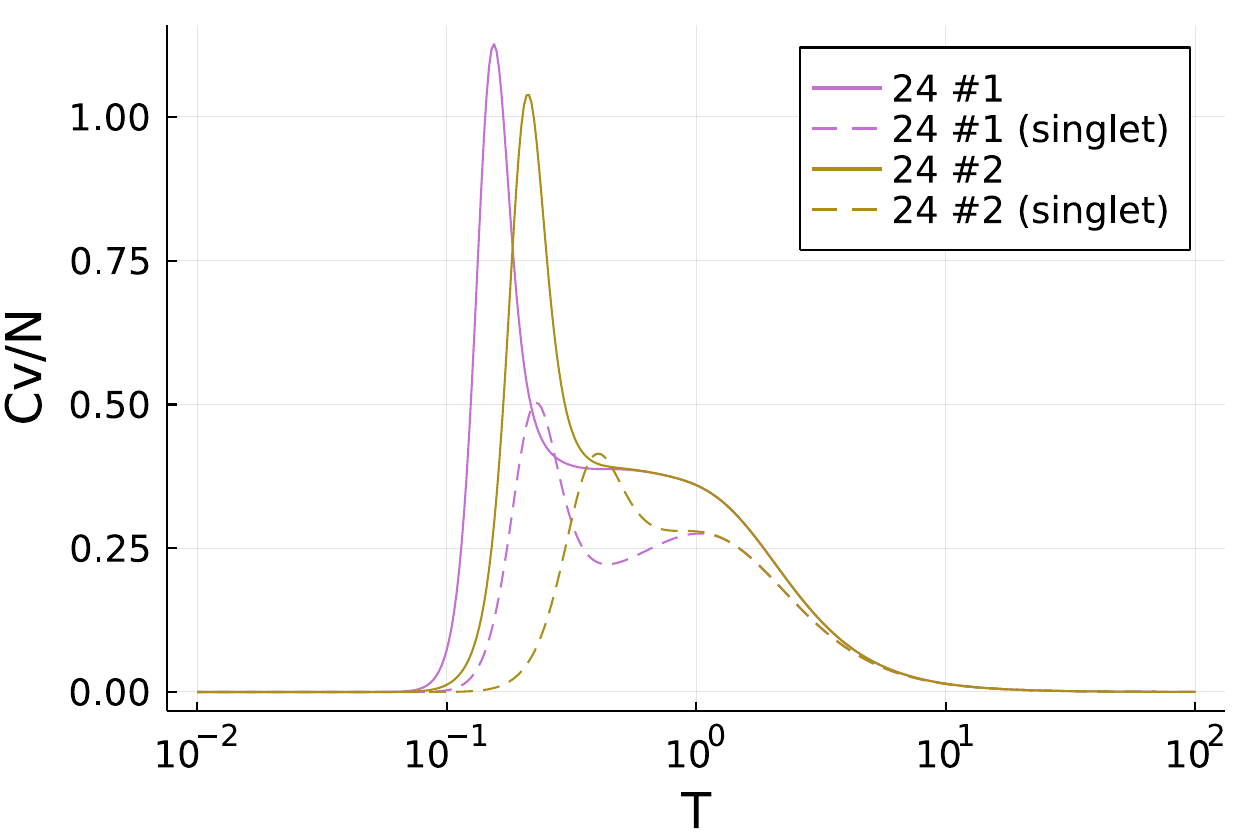}
    \caption{Comparison of specific heats between the whole symmetry sectors
    and the singlet sector.  Solid lines are the same as those in
    Fig.~\ref{main}(e) with $N=24$.  Dashed lines are corresponding
    specific heats calculated by confining the symmetry sector to
    the singlet sector.  It is clear that the low-temperature peak
    does not originate from singlet excitations.}
    \label{singlet}
\end{figure}

This is in contrast to what has been observed for the kagome spin liquid.
In the kagome case, the low-temperature physics below the spin gap is
dominated by neutral excitations~\cite{Lauchli2019}, which is consistent
with a recent expectation that the low-temperature effective theory consists
of neutral Dirac cones~\cite{He2017}.
On the other hand, in the present case, the low-temperature peak structure
is not dominated by singlet excitations, which is demonstrated by
the comparison with the specific heat restricted to the singlet subspace
(see Fig.~\ref{singlet}).
This is also consistent with the exact diagonalization observation that
the color gap $0.8882$ is much smaller than the singlet gap $1.1702$
for $N=24$.

We believe that our results are consistent with a gapped spin liquid
ground state, most probably a $Z_4$ spin liquid, rather than a gapless
one proposed previously~\cite{Corboz2012}.  This is because the
low-temperature peak associated with the $O(1)$ color gap seems
to be converged to the thermodynamic limit with the peak temperature
almost unchanged.  Indeed, the peak temperature is almost constant
for the 16b-, 20-, and 24-site calculations, and we cannot expect
that this peak disappears or moves drastically in the thermodynamic limit.
Of course, we cannot rule out a possibility that singlet excitations begin
to dominate below the color gap for $N>24$, but we believe that
this is an unlikely scenario.  In order to explain everything
observed in our calculations, the gapped spin liquid scenario
is more likely.  We note that previous gapless Dirac spin liquid
scenario~\cite{Corboz2012} has a problem of a relevant monopole
perturbation~\cite{Calvera2021} and also cannot explain the existence
of a color gap.  Based on this observation, we will discuss what
can be said from the assumption that the ground state is gapped
and symmetric from now on.

\textit{Rigorous proof of the Lieb-Schultz-Mattis-Affleck-Yamada-Oshikawa-Jackeli theorem}.---
Here we briefly give a rigorous version of the two-dimensional case of
the Lieb-Schultz-Mattis-Affleck-Yamada-Oshikawa-Jackeli (LSMAYOJ)
theorem~\cite{Lieb1961,Affleck1986,Yamada2018,Yamada2021su4},
which was first ``proven'' physically by the flux-insertion argument by
Yamada, Oshikawa, and Jackeli~\cite{Yamada2018,Yamada2021su4}.
This is a $G=\mathrm{PSU}(N)$ generalization of the Lieb-Schultz-Mattis theorem,
which was proven rigorously for one-dimensional systems by Ogata, Tachikawa, and
Tasaki~\cite{Ogata2021}.  Its generalization to two dimensions
is indeed straightforward.  We here give a very simple proof, utilizing the fact
that the Ogata-Tachikawa-Tasaki proof~\cite{Ogata2021} is directly treating
the infinite system.

Let us consider an infinite cylinder geometry with a spiral boundary condition
(sometimes called tilted or twisted boundary condition)~\cite{Yao2020,Yao2021,Nakamura2021}.
The spiral boundary condition is defined as follows.  First, let us begin with
$\mathbb{Z}^2$.  We glue it into an infinite cylinder by identifying
$(i,\,j) \in \mathbb{Z}^2$ with $(i + 1,\,j + L_y) \in \mathbb{Z}^2$,
where $L_y$ is a circumference.  The point is that this infinite cylinder
has a one-unit-cell translation symmetry as a one-dimensional system,
which we call a spiral translation symmetry.  Thus, for any $L_y$ we can
use the Ogata-Tachikawa-Tasaki theorem~\cite{Ogata2021}
to prove that the ground state is degenerate~\footnote{Here ``degenerate'' means
``impossible to have a unique gapped ground state''.} when the unit cell contains
a projective representation of $G$ and the interaction is short-ranged.
The degeneracy is maintained for any finite $L_y$ with a spiral boundary
condition, so the degeneracy will be preserved in the thermodynamic limit.

The advantage of this proof is that it is applicable to the discrete group $G$
case.  However, one caveat is that this kind of proof cannot give information on
the number of degeneracy.  The information on the degeneracy can be deduced
from the argument of the next section.

\textit{Physical proof of the existence of symmetry fractionalization}.---
The extension of the Lieb-Schultz-Mattis theorem not only can prove the ground state
degeneracy, but also can prove the existence of symmetry fractionalization of $G$.
Here we assume that $G$ is continuous and connected.  This was physically proven
in Refs.~\cite{Zaletel2015,Cheng2016} in the $G=\mathrm{SO}(3)$ case.
For simplicity, we also give our refined proof of this claim for $G=\mathrm{SO}(3)$
in the following.  This is actually achieved by proving the degenerate
ground states guaranteed by the LSMAYOJ theorem include different
one-dimensional symmetry-protected topological (SPT) states.

Assuming the unit cell containing a projective representation of $G$ and
the same setting as that of the proof of the LSMAYOJ theorem, a unique
and gapped ground state is forbidden.  Let us assume that ground states
are degenerate with a gap.  The point is that the SPT phase is distinguished
by the Ogata index $\sigma_x^\mathrm{R} \in H^2(G,\mathrm{U}(1))=Z_2$,
where $x = (i,\,j)$ is the origin of the unit cell~\cite{OgataTasaki2019,Ogata2020,Ogata2019,Ogata2021}, and
$\sigma_x^\mathrm{R} = 0$ and $\sigma_x^\mathrm{R} = 1$ are essentially equivalent.
This is because $\sigma_x^\mathrm{R} = 0$ and $\sigma_x^\mathrm{R} = 1$ are exchanged
just by changing the definition of the origin $x$ through a single spiral
translation.  This ``democracy'' of the Ogata index automatically
guarantees that the same number of $\sigma_x^\mathrm{R} = 0$ states and
$\sigma_x^\mathrm{R} = 1$ states should be included in the degenerate ground
states.  Such a distinction must be independent of the way of the
dimensional reduction, which physically finishes the
proof~\footnote{The meaning of degeneracy is different between
the infinite cylinder and the thermodynamic limit.  On the infinite
cylinder, the degeneracy between the $\sigma_x^\mathrm{R} = 0$ and
$\sigma_x^\mathrm{R} = 1$ states usually means the translation
symmetry breaking.  However, this symmetry breaking is an artifact
of the spiral boundary condition imposed on the infinite cylinder, and
the symmetry is restored in the two-dimensional limit $L_y \to \infty$
for the topologically ordered case.}.

It is not clear that the degeneracy of different SPT states means the
symmetry fractionalization.  Intuitively, this can be understood as follows.
On the finite cylinder with an even number of sites, the distinction of
SPT phases can be captured by edge states.  A trivial SPT phase
has no edge states, while a nontrivial SPT phase has spin-1/2 edge states.
In the large-cylinder limit, these two correspond to a vacuum sector
and a spinon sector of the topological phase, respectively,
assuming that the vacuum sector does not have an edge state.
This means that an original spin-1 excitation (magnon) is fractionalized
into two spinons and separated into both edges with long-range entanglement.
This is nothing but the proof of the existence of fractionalization
in the $G=\mathrm{SO}(3)$ case.

The generalization to the case with $G=\mathrm{PSU}(N)$ is straightforward.
In the case of the present $\mathrm{SU}(4)$ model, $G=\mathrm{PSU}(4)$ and
$H^2(G,\mathrm{U}(1)) = Z_4$.  As for the honeycomb lattice model,
$i$ and $i+2$ ($i=0,\, 1$) in $H^2(G,\mathrm{U}(1)) = Z_4$ are democratic,
so this proves the degeneracy of $\sigma_x^\mathrm{R} = i$ and $\sigma_x^\mathrm{R} = i+2$
states as ground states.

In the thermodynamic limit, this not only means the existence of topological
order, but also the fractionalization of the symmetry $G$, \textit{i.e.}
the existence of spinons and orbitalons which behave as a (6-dimensional)
projective representation of $G$.  On the finite cylinder, the vacuum
sector without edge anyons, and the anyon sector with edge anyons must
be almost degenerate.

\textit{Counterexamples}.---
There exists a counterexample like Wen's plaquette model~\cite{Wen2003}
for the discrete group $G$ case.  This is because degenerate ground
states guaranteed by the Lieb-Schultz-Mattis theorem are exchanged directly
by $G$, not by translation.  The Ogata index is defined only when all
the ground states in the infinite system are $G$-symmetric, and thus the
above discussion fails in this case.  However, this never happens when $G$
is continuous and connected, so our conclusion is not affected by such
a possibility.  More detailed discussions can be found in
Refs.~\cite{Zaletel2015,Cheng2016}.

\textit{Origin of the peak}.---
So far we have identified the energy scale of the specific heat peak as
the color gap, but the physical origin of an existing peak is still
not clear.  If we assume that there is no symmetry breaking, we have
to come up with another mechanism beyond the Landau theory to describe
the existence of a peak.

In contrast to the Landau theory, fermionic partons can acquire
a gap without spontaneous symmetry breaking because a four-fermion
condensate dynamically generates a gap without a bilinear term.
This condensate can be described in terms of fermionic partons
(spinons and orbitalons) $f_{ia}$, where $i$ stands
for a site index and $a = 1$,\dots,4 stands for a color index,
as follows~\cite{Calvera2021,Zhang2020}.
\begin{equation}
    \Delta = \langle \epsilon_{abcd} f_{ia} f_{ib} f_{ic} f_{id}\rangle,
\end{equation}
where $\epsilon_{abcd}$ is a completely antisymmetric tensor.  Thus,
we would see a crossover between the $\Delta \sim 0$ disordered phase
to the $\Delta > 0$ ordered phase at some temperature $T_c$, which
potentially produces a peak-like behavior in the specific heat.

The gap opening and the symmetry fractionalization occurs
simultaneously at $T_c$ because we have proven that the symmetric
gapped state inevitably leads to symmetry fractionalization.
In this sense, we can regard the low-temperature peak as
reminiscence of fractionalization, while it is not clear how
the action of $\mathrm{SU}(4)$ is transformed in this crossover.

\textit{Discussion}.---
The existence of a low-temperature peak implies that the ground state of
the $\mathrm{SU}(4)$ Heisenberg model on the honeycomb lattice
is a gapped spin-orbital liquid rather than a gapless liquid proposed
previously~\cite{Corboz2012}.
If the ground state is really gapped and symmetric, the LSMAYOJ theorem
guarantees the ground state degeneracy, \textit{i.e.} the existence of topological order.
Even more, the extended version of this theorem proves the fractionalization of
the $\mathrm{PSU}(4)$ symmetry, and the existence of spinons and orbitalons
which behave as a projective representation of $\mathrm{PSU}(4)$.

Finally, we conjecture a generalized version of the Lieb-Schultz-Mattis-type
theorem.  The claim is that if the representation of $G$ per unit cell is
projective, there must be multiple degenerate ground states which are
distinguished as one-dimensional $G$-SPT phases after the dimensional reduction.
This version is yet to be proven mathematically rigorously, and left for
the future work.

\begin{acknowledgments}
We thank T.~Mizushima, F.~Pollmann, T.~Shimokawa, Y.~Tada, K.~Totsuka, H.~Ueda, and H.~Yoshida.
This work was supported by JSPS KAKENHI Grant Nos. JP21H01039 and JP22K14005,
and by JST CREST Grant Number JPMJCR19T5, Japan.
M.G.Y. is supported by Multidisciplinary Research Laboratory System for Future Developments,
Osaka University.
The computation in this work has been done using the facilities of the Supercomputer Center,
the Institute for Solid State Physics, the University of Tokyo, and partly
using large-scale computer systems at the Cybermedia Center, Osaka University.
\end{acknowledgments}

\bibliography{paper}

\end{document}